\documentclass[twocolumn]{aastex631}

\usepackage{multirow} 
\usepackage{amsmath}
\usepackage{enumitem}
\usepackage{subfigure}
\usepackage{xcolor}

\begin{document}
\title{Lopsided and Bulging Distribution of Satellites around Paired Halos. I. Observational Measurements and Comparison with Halo-based Models}

\correspondingauthor{Yanhan Guo}
\email{guoyh21@mails.tsinghua.edu.cn}
\correspondingauthor{Cheng Li}
\email{cli2015@tsinghua.edu.cn}

\author{Yanhan Guo}
\affiliation{Department of Astronomy, Tsinghua University, Beijing 100084, China}

\author{Qinglin Ma}
\affiliation{Department of Astronomy, Tsinghua University, Beijing 100084, China}

\author[0000-0002-8711-8970]{Cheng Li}
\affiliation{Department of Astronomy, Tsinghua University, Beijing 100084, China}



\begin{abstract}
We investigate the angular distribution of satellite galaxies in and around pairs of galaxy groups in SDSS and compare the results with the satellite distribution in pairs of dark matter halos by constructing mock catalogs that account for the same selection effects as the observational sample. We find that the angular distribution of satellites in both SDSS and the mock catalog exhibits a pronounced tendency towards lopsidedness, with satellites preferentially located between the two central galaxies. Additionally, there is a significant bulging distribution characterized by a higher concentration of satellites along the line connecting the two centrals compared to those found perpendicular to it. The lopsided and bulging distributions strengthen as pair separation and halo mass increase. The mock catalog successfully reproduces the observational results across all cases considered. We find that the lopsided and bulging distribution of satellites can largely be explained by overlapping two randomly selected halos matched in mass to the actual halo pairs, along with their surrounding satellite distribution, provided that the alignment between the orientations of the halos and the line connecting the halo pairs is taken into account. This suggests that the angular distribution of satellites is a natural consequence of the formation
and evolution of large-scale structure in a $\Lambda$CDM universe, eliminating the need to introduce other physical origins.
\end{abstract}



\section{Introduction} \label{sec:intro}

In a $\Lambda$ cold dark matter ($\Lambda$CDM) universe, galaxies form through the cooling and condensation of gas at the centers of dark matter halos, which form and grow hierarchically through gravitational instability and mergers (e.g. \citealt{White1978,Davis1985,Bond1996,Mo2010}). In this framework, the distribution of galaxies is theoretically expected and has been observationally confirmed to closely follow the distribution of dark matter halos across a broad range of spatial scale~\cite[e.g.][]{Davis1985}. Furthermore, numerical simulations and observations have established that dark matter halos, which manifest as triaxial ellipsoids \citep[e.g.][]{Jing2002}, are not oriented at random; instead, they exhibit spatial alignment with the orientation of central galaxies (although with a median misalignment angle of $\sim35^\circ$; e.g. \citealt{Wang2008,Okumura2009,Niederste2010,Zhang2021}), as well as with the orientation and spatial distribution of satellite galaxies \citep{Brainerd2005,Yang2006,Faltenbacher2007,Faltenbacher2008,Georgiou2019}, and the large-scale distribution of halos and galaxies \citep[e.g.][]{Faltenbacher2009,Li2013,Zhang2013,Chen2015}. These various alignment phenomena can be naturally attributed to the large-scale tidal field and the preferred accretion of matter along filamentary structures. The alignment of galaxies in the local Universe, observed over scales ranging from galaxy sizes to more than $100$ Mpc, can be effectively reproduced by simple halo-based models of galaxy distribution, provided that the misalignment between the orientations of halos and central galaxies is taken into account \citep[e.g.][]{Li2013}.

Recently, \citet{Brainerd2020} discovered that satellites around {\em isolated} luminous galaxies in the Sloan Digital Sky Survey (SDSS; \citealt{York2000}) tend to be located predominantly on one side of their host galaxies. This ``lopsided'' distribution of satellites has also been observed in hydrodynamic simulations, such as Illustris-TNG~\citep{Wang2021,Mangis2024}, as well as in mock catalogs constructed from the Millennium simulation~\citep{Samuels2023}. The findings indicate that stronger lopsidedness is associated with satellite galaxies around hosts with lower masses and bluer colors, in addition to more massive neighboring galaxies, greater distances from the hosts, and lower redshifts. Moreover, lopsided distributions of galaxies have been identified in galaxy clusters~\citep{Mangis2024} and among nearby dwarf galaxy systems~\citep{Heesters2024,Martinez2024}. Notably, \citet{Heesters2024} observed such lopsided distributions around red early-type host galaxies in the MATLAS and ELVES surveys. More recently, \citet{Liu2024} traced back the satellite systems in simulations to high redshifts, attributing the lopsidedness to recent anisotropic accretion of subhalos.

The lopsided distribution of satellites (or subhalos) has also been identified around galaxy pairs (halo pairs), marked by an overabundance of satellites/subhalos in the region {\em between} the paired hosts compared to the outer regions. Moreover, these satellites/subhalos tend to align along the axis connecting the pair, giving the system a bulge-like shape. For instance, within the Local Group, dwarf satellites are found to preferentially occupy the space between the Milky Way (MW) and M31 \citep{McConnachie2006,Conn2013}. Utilizing the SDSS galaxy sample, \cite{Libeskind2016} found similar lopsided and bulging tendencies in the satellite distribution around galaxy pairs selected as analogs of the MW-M31 system. Furthermore, the lopsided and bulging signals were found to be weaker for galaxy pairs residing in filaments compared to those located outside filaments, suggesting that the signals arise from local gravitational effects rather than the filamentary distribution and large-scale accretion of matter \citep{Libeskind2016}. An analysis of $\Lambda$CDM simulations by \cite{Pawlowski2017} revealed similar but more pronounced lopsided and bulging distributions, attributing the weaker signal observed by \cite{Libeskind2016} to background and foreground contamination in the data. By tracing the orbital trajectories of satellites around halo pairs in simulations, \citet{Gong2019} demonstrated that the lopsided and bulging distributions are primarily contributed by satellites approaching the paired hosts for the first time; thus, the anisotropic signature diminishes once satellites have passed by.

Considering that galaxy pairs exhibit significant alignment with their host filaments \citep{Tempel2015,Epps2017} alongside various forms of spatial alignment and the lopsided distribution of satellites around isolated halos and galaxies, the lopsidedness and bulging signals around halo/galaxy pairs may arise simply from the overlap of two halos, each associated with somewhat aligned and lopsided satellite distributions. Previous studies have tested this hypothesis by conducting similar analyses on samples of ``overlapping'' halos or galaxies, generated by artificially placing two isolated halos or galaxies along with their subhalos or satellites at separations matching those of real pairs. These tests generally yielded no or weak signals, seemingly ruling out the overlap effect as the origin of the lopsidedness and bulging distributions of satellites around galaxy pairs. However, in these tests, the two isolated halos/galaxies were randomly oriented, neglecting the spatial alignment of their orientations concerning the surrounding satellites. It is crucial to appropriately incorporate the alignment signal in assessing the overlap effect before arriving at a conclusive judgment regarding its contribution to real pairs of halos or galaxies.

In this paper, we extend the exploration of lopsided and bulging distributions of satellites around galaxy pairs, which has previously focused primarily on the MW-M31 system and its analogs, by selecting a large sample of paired central galaxies from SDSS and investigating the dependence of lopsidedness and bulging signals on host halo mass, halo mass ratio, and pair separation. In particular, we carefully account for the spatial alignment of galaxy orientations in relation to the large-scale structure when testing the overlap effect. Additionally, we utilize numerical simulations of dark matter halos and construct mock catalogs that replicate the same selection effects present in the real galaxy sample. We perform analogous analyses on the mock catalogs and compare the results with observational measurements. As we will demonstrate, once alignment is considered in the overlap effect, both the lopsidedness and bulging signals can be effectively explained, with the observational measurements of these signals being well reproduced by the numerical simulations. Combined with previous studies on galaxy alignment and the lopsided distribution of satellites around isolated galaxies, our results illustrate that the anisotropic distribution around galaxy pairs is a natural consequence of the formation and evolution of large-scale structure in a $\Lambda$CDM universe. In a parallel paper (\citealt{ma2025}), we measure the satellite distribution around halo pairs in both three dimensions and two-dimensional planes, and make comparisons with those from the mock catalog as done in this paper, in order to understand the effect of projection and observational sample selection effects. 

This paper is organized as follows. In~\autoref{sec:2}, we introduce the criteria for sample selection and present the measurements of the anisotropic distribution in the observational sample. In~\autoref{sec:mock}, we conduct a similar analysis with mock catalogs and compare the results with observational measurements. In~\autoref{sec:overlap}, we describe the construction of our anisotropic overlap sample and explore the origin of the anisotropic distribution of satellites. Finally, we summarize our results in~\autoref{sec:sum}. Throughout the paper, we adopt a $\Lambda$CDM cosmology with $\Omega_{m} = 0.3089$ and $h = 0.6774$, following  \citet{Planck2016}.

\section{Observational Measurements} \label{sec:2}


\subsection{Data} \label{subsec:data}

The observational data used in this work are drawn from the New York University Value Added Galaxy Catalog (NYU-VAGC)\footnote{\url{http://sdss.physics.nyu.edu/vagc/}}, which was constructed by \cite{Blanton2005} based on the final data release of the SDSS \citep{SDSS-DR7}. The sample comprises about half a million galaxies with $r$-band Petrosian apparent magnitudes $r<17.6$ and spectroscopically measured redshifts in the range $0.01<z<0.2$. Stellar masses of the galaxies are estimated by \citet{Blanton-Roweis-2007} using the Petrosian magnitudes across the SDSS $u$, $g$, $r$, $i$ and $z$ bands, assuming a stellar initial mass function from \citet{Chabrier-2003}. 

We begin by identifying central galaxies from the full sample using the following approach. First, we assume each galaxy to be the central galaxy of its associated dark matter halo and estimate a halo mass ($M_{\rm{halo}}$) based on the stellar-to-halo mass relation derived by \cite{Guo2010}. Next, for a given galaxy, we locate all companion galaxies within a projected radius of $r_{p} < 2.5 r_{\text{vir}}$ and a line-of-sight velocity separation of $\Delta v < 1000 ~ \mathrm{km/s}$. The galaxy is classified as a central galaxy if it is the most massive among all companion galaxies within this volume. Here, the virial radius $r_{\text{vir}}$ is computed from $M_{\rm{halo}}$ and is defined as the radius within which the mean matter density is 200 times the critical density of the universe. The choice of a projected radius of $2.5 r_{\text{vir}}$ aims to optimize both the completeness and purity of the resulting central galaxy sample, as determined by tests conducted on the mock catalog that we construct in the next section.  

We subsequently identify pairs of central galaxies (and thus halo pairs) from the central galaxy sample using the following criteria: (i) dark matter halo mass in the range $10^{11}~h^{-1} \mathrm{M_{\odot}} < M_{\text{halo}} < 10^{15}~h^{-1} \mathrm{M_{\odot}}$, and (ii) a projected separation between the two central galaxies in the range $0.25~h^{-1} \mathrm{Mpc} < d_{\text{sep}} < 2.0~h^{-1} \mathrm{Mpc}$, along with a line-of-sight velocity separation of $\Delta v_{\text{pair}} < 1000~\mathrm{km/s}$. In this context, the more massive halo is referred to as the ``primary halo'', while the less massive halo is termed the ``secondary halo''. Moreover, we require that there be no third halo whose mass exceeds half of the mass of the secondary halo and whose center lies within a projected distance of $d_{\text{sep}}$ from the midpoint of the line connecting the two central galaxies, as well as within a line-of-sight separation of $1000~\mathrm{km/s}$ from the mean velocity of the two central galaxies. It is important to note that these criteria are similar to those employed in \citet{Gong2019}, which aimed to select analogs of the MW-M31 system. However, we utilize wider ranges for both $M_{\text{halo}}$ and $d_{\text{sep}}$ to expand the analysis to halo pairs of varying masses, mass ratios, and separations. 

Following \citet{Gong2019}, for each halo pair, we define ``satellites'' as all galaxies within a projected separation $r_{p} < 0.5 d_{\text{sep}}$ and a line-of-sight velocity separation $\Delta v < 1000~\mathrm{km/s}$ from the central galaxy of their nearest host halo. This definition intentionally differs from the conventional use of ``satellite", which refers to non-central galaxies located within the virial radius of their host dark matter halo. For halo pairs separated by more than the primary halo's virial radius, our sample includes not only genuine satellites but also interloping galaxies from the surrounding environment. Conversely, for pairs separated by less than the primary halo's virial radius, our definition captures only a subset of the actual satellites. This operational definition allows us to measure the overall galaxy distribution surrounding the halo pair across a wide range of spatial scales, rather than distributions strictly localized to individual halos. Consequently, it enables us to explore the combined effects of both large-scale structure and local gravitational interactions. As shown below, our results---largely explained by the alignment of halo orientations with their surrounding large-scale structure---could not have been revealed using conventionally-defined satellites.

These criteria yield a total of 180,390 halo pairs and 191,760 satellites.

\begin{figure}
  \centering
  \includegraphics[width=1\linewidth]{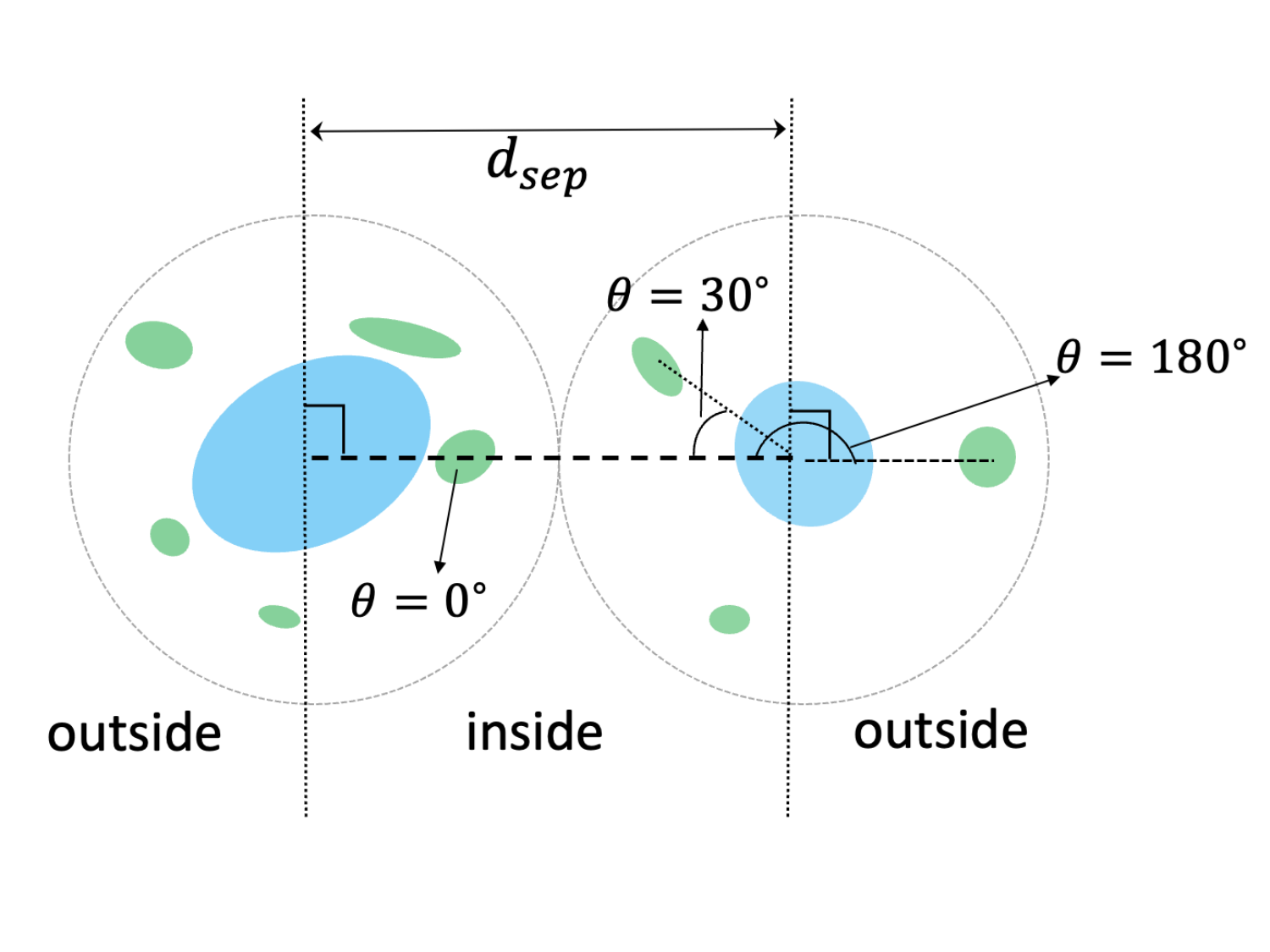}
  \caption{The schematic of a typical halo pair, as observed on the sky. The two blue ellipses represent the paired halos, which are central galaxies selected from the SDSS sample and separated by a distance of  $d_{\text{sep}}$. Galaxies with projected separation $d_{\text{sat}}<0.5d_{\text{sep}}$ from their nearest host are identified as ``satellites" and are plotted as green ellipticals. The position angle of the satellites, $\theta$, is measured between the line connecting it to its host central galaxy and the line connecting the two centrals. A satellite is located ``inside'' the halo pair if $0^\circ\leq\theta\leq 90^\circ$, or ``outside'' the halo pair if $\theta>90^\circ$.}
  \label{fig:halo_pair_diagram}
\end{figure}

\begin{figure*}[ht!]
  \centering
  \includegraphics[width=0.9\linewidth]{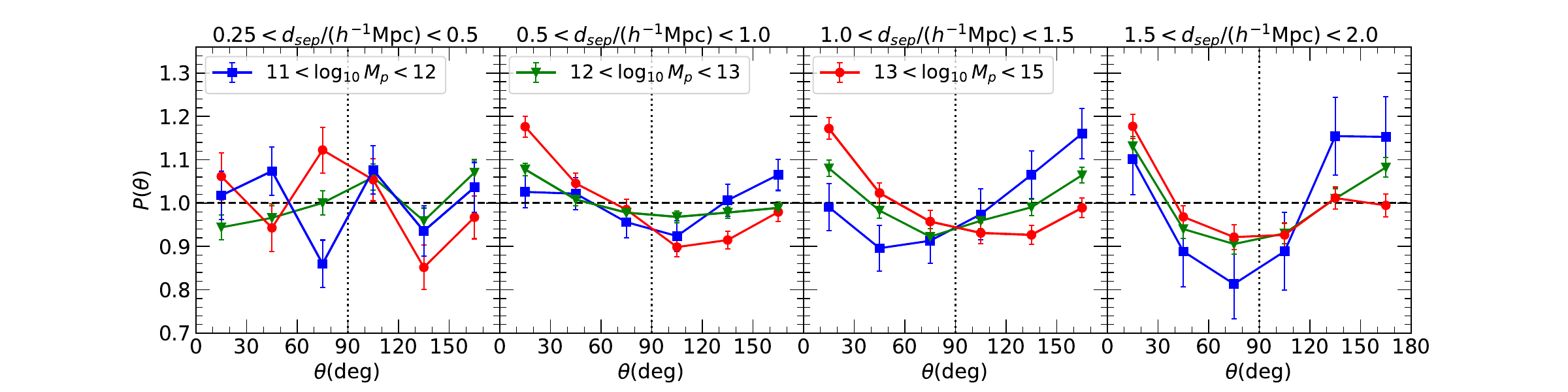}
  \includegraphics[width=0.9\linewidth]{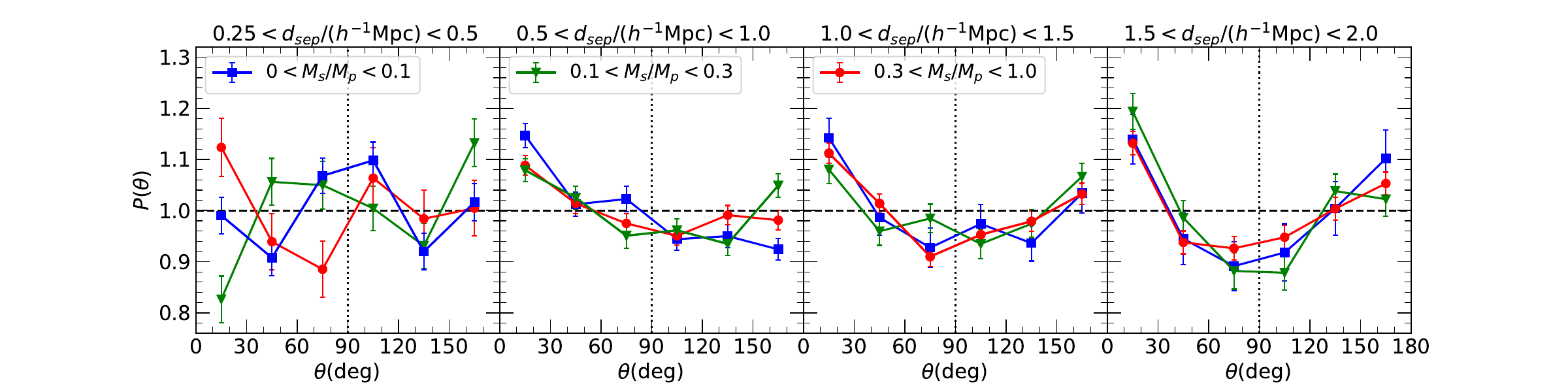}
  \caption{Angular distribution of satellites around halo pairs from SDSS, $P(\theta)$, as measured for pairs of varying separation, primary halo mass and secondary-to-primary halo mass ratio, as indicated.}
  \label{fig:ang_sdss}
\end{figure*}

\begin{figure*}[ht!]
  \centering
  \includegraphics[width=0.9\linewidth]{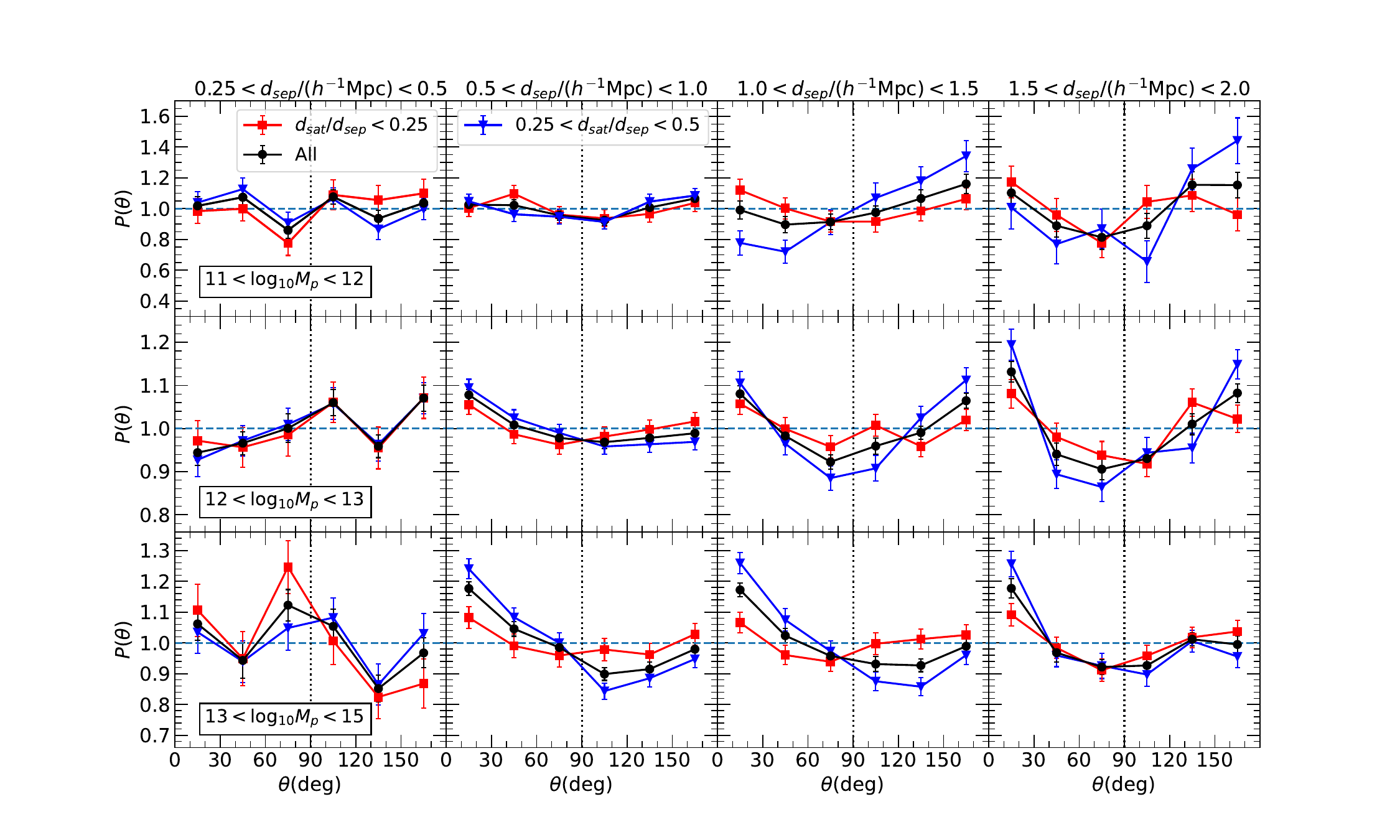}  
  \caption{Angular distribution of satellites around halo pairs from SDSS, $P(\theta)$, as measured for pairs of varying separation, primary halo mass and satellite-to-host distance, as indicated.}
  \label{fig:Ang_sdss_dsat}
\end{figure*}

\subsection{Satellite Distribution} \label{subsec:mass}

We measure the distribution of satellite galaxies in relation to the orientation of halo pairs. As illustrated in \autoref{fig:halo_pair_diagram}, we define a position angle $\theta$ on the sky for each satellite as the angle between the line connecting the satellite to its host central galaxy (the nearer of the two centrals) and the line connecting the two central galaxies. With this definition, $\theta$ ranges from $0^{\circ}$ to $180^{\circ}$. Angles $\theta < 90^{\circ}$ correspond to satellites located ``inside'' the halo pair (between the two central galaxies), while $\theta > 90^{\circ}$ indicates satellites that are ``outside'' the halo pair. Satellites positioned along the line connecting the two centrals have $\theta = 0^{\circ}$ if they are inside the halo pair, or $\theta = 180^{\circ}$ if they are outside. Lopsidedness of the satellite distribution is quantified by comparing the abundance of satellites located inside ($\theta<90^{\circ}$) and outside ($\theta>90^{\circ}$) their halo pairs, while the bulging distribution is assessed by comparing the satellite abundance along the line connecting the two central galaxies with those perpendicular to the connecting line ($\theta < 45^{\circ}$ versus $45^{\circ} < \theta < 90^{\circ}$ for satellites inside the halo pairs, and $\theta > 135^{\circ}$ versus $90^{\circ} < \theta < 135^{\circ}$ for those outside the halo pairs).

\autoref{fig:ang_sdss} displays the distributions of satellite position angle, $P(\theta)$, for halo pairs with different separations (panels from left to right), primary halo masses (upper panels), and secondary-to-primary halo mass ratios (lower panels). To estimate the errors of $P(\theta)$, we randomly shuffle the position angles of central galaxies and recalculate the satellite position angle distribution. This process is repeated 500 times\footnote{Our test shows that 500 random samples ensures stable error estimates for $P(\theta)$, with fluctuations reduced to $\lesssim 5\%$.}, and the errors of $P(\theta)$ are estimated by the scatter between the random samples. When calculating $P(\theta)$, we have corrected for the effect of volume incompleteness in the SDSS galaxy sample by weighting each pair by the inverse of the volume in which both the host centrals and the satellites can be included in the sample. In the parallel paper (\citealt{ma2025}), our test based on mock catalogs shows that this correction scheme works very well, resulting in $P(\theta)$ measurements that agree well with the measurements obtained directly from the simulation. In addition, we have tested on potential biases caused by incompleteness at survey edges, due to the missing satellites of halo pairs that are located towards the boundaries of either survey footprint or redshift range. We found our results presented in the rest of this paper remain unchanged when those halo pairs near the survey edges are removed from the sample, demonstrating that our results are not affected by the edge effects.

As can be seen, for all halo pairs considered, except those with the smallest separations ($d_{\text{sep}} < 0.5~h^{-1} \mathrm{Mpc}$), the satellite position angles exhibit a lopsided distribution, with unequal $P(\theta)$ values between $\theta < 90^{\circ}$ and $\theta > 90^{\circ}$. Furthermore, in most cases, $P(\theta)$ displays a bulging signal, with a minimum at an intermediate $\theta$, indicating that the satellites are preferentially aligned along the connection line between the halo centers. Noticeably, the angle at which the minimum occurs increases with primary halo mass for pairs of given separation, but it appears to decrease with pair separation for pairs of given halo mass. Both the lopsidedness and bulging signals increase as the separation between pairs increases. 

Halo pairs with high masses ($M_{\text{p}} > 10^{13} M_\odot$) show strong lopsidedness towards small angles and pronounced bulging signals at $\theta < 90^{\circ}$, with this trend being only weakly dependent on pair separation. This result suggests that satellites in high-mass halos prefer to be located between the two halo centers and aligned with the orientation of the halo pair. In contrast, for low-mass halo pairs, while satellites also tend to align with the pair orientation, they are preferentially found outside rather than inside the halo pairs. Additionally, a strong dependence on pair separation is observed for low-mass halo pairs, with stronger bulging signals but weaker lopsidedness as the pair separation increases. Finally, when the pair separation is fixed, we find no obvious dependence of $P(\theta)$ on the secondary-to-primary halo mass ratio, as indicated in the lower panels.

In \autoref{fig:Ang_sdss_dsat}, we further examine the dependence of $P(\theta)$ on the distance of satellites to their host central galaxy. We scale the satellite-to-host distance $d_{\text{sat}}$ by the pair separation $d_{\text{sep}}$ and divide all satellites into two subsamples based on the ratio of $d_{\text{sat}}/d_{\text{sep}}$. Generally, for halo pairs with a given pair separation and primary halo mass, the distribution of satellites located further from their halo centers is observed to be more bulging and lopsided compared to those closer to the halo centers. This effect becomes more pronounced for more massive and more widely separated halo pairs. This finding aligns with the suggestion made by \citet{Gong2019}, which proposed that the anisotropic distribution arises from dynamically young accretion events. Our results support their hypothesis, as recently accreted galaxies are typically found at greater distances from their host halos.

\begin{figure*}[ht!]
  \centering
  \includegraphics[width=0.9\linewidth]{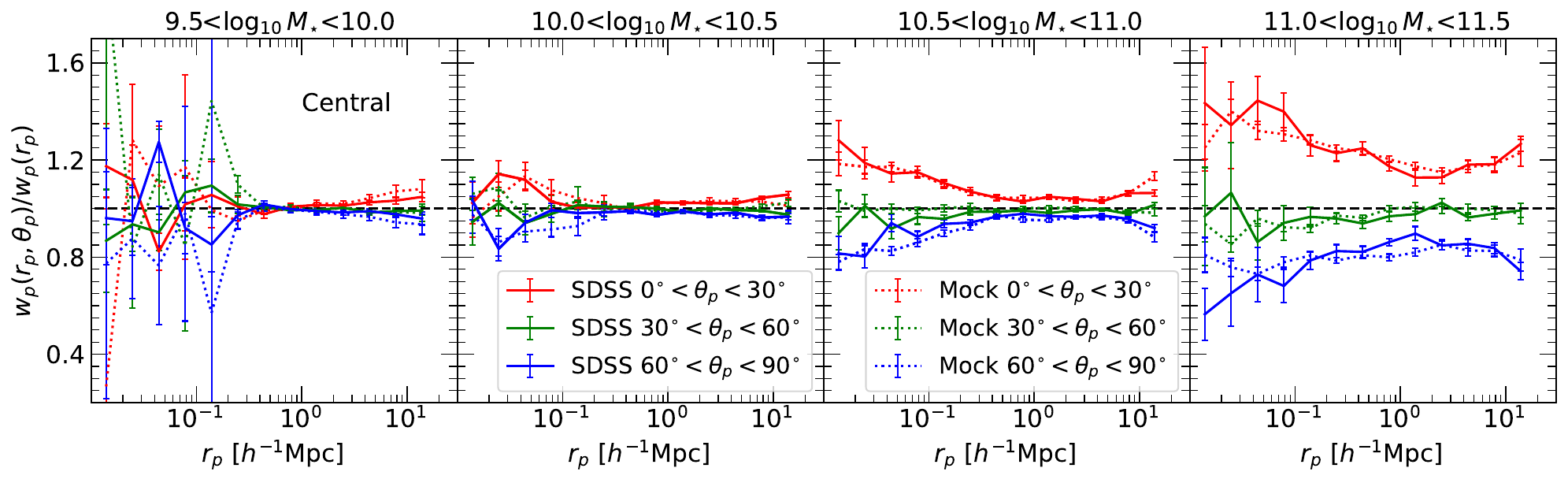}
  \caption{Ratios of the alignment cross-correlation function $w_p(r_p,\theta_p)$ at three non-overlapping intervals of $\theta_p$ relative to the conventional cross-correlation function $w_p(r_p)$, as measured for galaxies of different stellar mass, as indicated. Solid and dotted lines in each panel represent the SDSS sample and the mock catalog, respectively.}
  \label{fig:test_mock_theta}
\end{figure*}

\section{Comparison with halo-based models} \label{sec:mock}

To make comparison with theoretical models, we first populate dark matter halos in a cosmological simulation with galaxies of varying properties, which are determined through an empirical galaxy formation model. Subsequently, we establish the orientations of our model galaxies and construct mock catalogs that replicate the same selection effects as the SDSS galaxy sample. Finally, we measure the satellite distribution in the mock catalog and compare the measurements with the SDSS-based measurements presented above.

\subsection{Simulation and galaxy model}

The simulation utilized in this study is {\it IllustrisTNG300}, part of The Next Generation Illustris dark matter only simulation (IllustrisTNG-DMO; \citealt{Marinacci2018,Naiman2018,Nelson2018,Pillepich2018,Springel2018}). This simulation contains $2500^3$ dark matter particles with a particle mass of $7.0 \times 10^{7}~h^{-1} \mathrm{M_{\odot}}$ within a periodic box of size $L = 205~h^{-1} \mathrm{Mpc}$ (approximately $300 ~\mathrm{Mpc}$) on each side. The adopted cosmological parameters are consistent with measurements from \cite{Planck2016}: $\Omega_{m} = 0.3089$, $\Omega_{b} = 0.0486$, and $h = 0.6774$. The simulation offers 100 snapshots corresponding to redshifts ranging from $z = 20$ to $z = 0$. Dark matter halos and subhalos are identified using the friends-of-friends algorithm (FOF; \citealt{Davis1985}) and the SUBFIND algorithm (\citealt{Springel2001}), respectively. In this study, we utilize the catalog of halos and subhalos at $z = 0$.

We assume that each subhalo hosts a galaxy, and we assign a stellar mass to each galaxy using the subhalo abundance matching (SHAM) model (\citealt{Vale2004, Conroy2006, Moster2010, Guo2010, Reddick2013}). In practice, we derive the relationship between galaxy stellar mass $M_{\star}$ and $v_{\text{peak}}$, which is the highest circular velocity that a halo has attained throughout its entire merger history. This is done by matching the number density of halos with $v_{\text{peak}}$ above a given threshold to the number density of galaxies with $M_{\star}$ above a corresponding threshold, as computed from the galaxy stellar mass function in the local Universe measured by \cite{Li2009}. Next, following \cite{Meng2024}, we further assign an $r$-band absolute magnitude $M_{\text{r}}$ to each galaxy based on the relationship between $M_{\star}$ and $M_{\text{r}}$ observed in SDSS galaxies. For further details, readers are referred to \cite{Meng2024}. 

\subsection{Construction of the mock catalog}
\label{sec:mock_construction}

We follow the method described in \cite{Li2006b} and \cite{Meng2024} to construct our mock catalog, which incorporates the same selection effects as the SDSS galaxy sample used in the previous section. Briefly, we construct a light cone using simulation snapshots to cover the same volume and redshift range as the real survey, and then incorporate all relevant observational selection effects, such as redshift-dependent incompleteness due to the limiting magnitude, sky position-dependent incompleteness, the k-correction effect, and others. More details can be found in \cite{Li2006b}.

For each galaxy, we determine its orientation using the dark matter particles within the central $0.04 R_{\text{vir}}$ of the host halo. This approach assumes that the stellar component of a galaxy can be effectively represented by the central region of its parent dark matter halo (cf. \citealt{Faltenbacher2009}). The choice of $0.04 R_{\text{vir}}$ is based on the consideration that the effective radius of disk galaxies may be approximated by the product of the halo spin $\lambda_{\mathrm{halo}}$ and the virial radius $R_{\text{vir}}$ (\citealt{Mo1998}), and that halo spins follow a log-normal distribution with a median value of $\langle \lambda_{\mathrm{halo}} \rangle = 0.03-0.05$ (\citealt{Bullock2001,Bett2007,Danovich2015}). Assuming that the shape of each galaxy can be modeled by a triaxial ellipsoid, we use the dark matter particles in the central region to calculate the weighted inertia tensor \citep{Allgood-2006}. The eigenvectors of this tensor are used to determine the three axes of the ellipsoid, including both their lengths and directions. The longest axis is taken as the major axis of the galaxy, and is projected onto the sky to indicate the {\it observed} major axis in the mock catalog.  

Previous studies have established that the orientations of central galaxies are not perfectly aligned with the orientations of their host halos. Instead, there exists a misalignment angle between their major axes, which follows a Gaussian distribution with a mean of zero and a width of $\sigma_{\theta}$ (\citealt{Wang2008,Okumura2009}). Accordingly, we incorporate a misalignment angle for each of the mock galaxies by adopting a Gaussian distribution with a mean of zero and a mass-dependent width $\sigma_{\theta}$, which is given by:
\begin{equation}
    \sigma_{\theta} = \left \{
    \begin{aligned}
        &-45(M_{\star}-10.5)^2+40,&\text{for halos}\\ 
        &-45(M_{\star}-10.5)+60,&\text{for subhalos}
    \end{aligned}
    \right.
\end{equation} 

\begin{figure*}[ht!]
  \centering
  \includegraphics[width=0.9\linewidth]{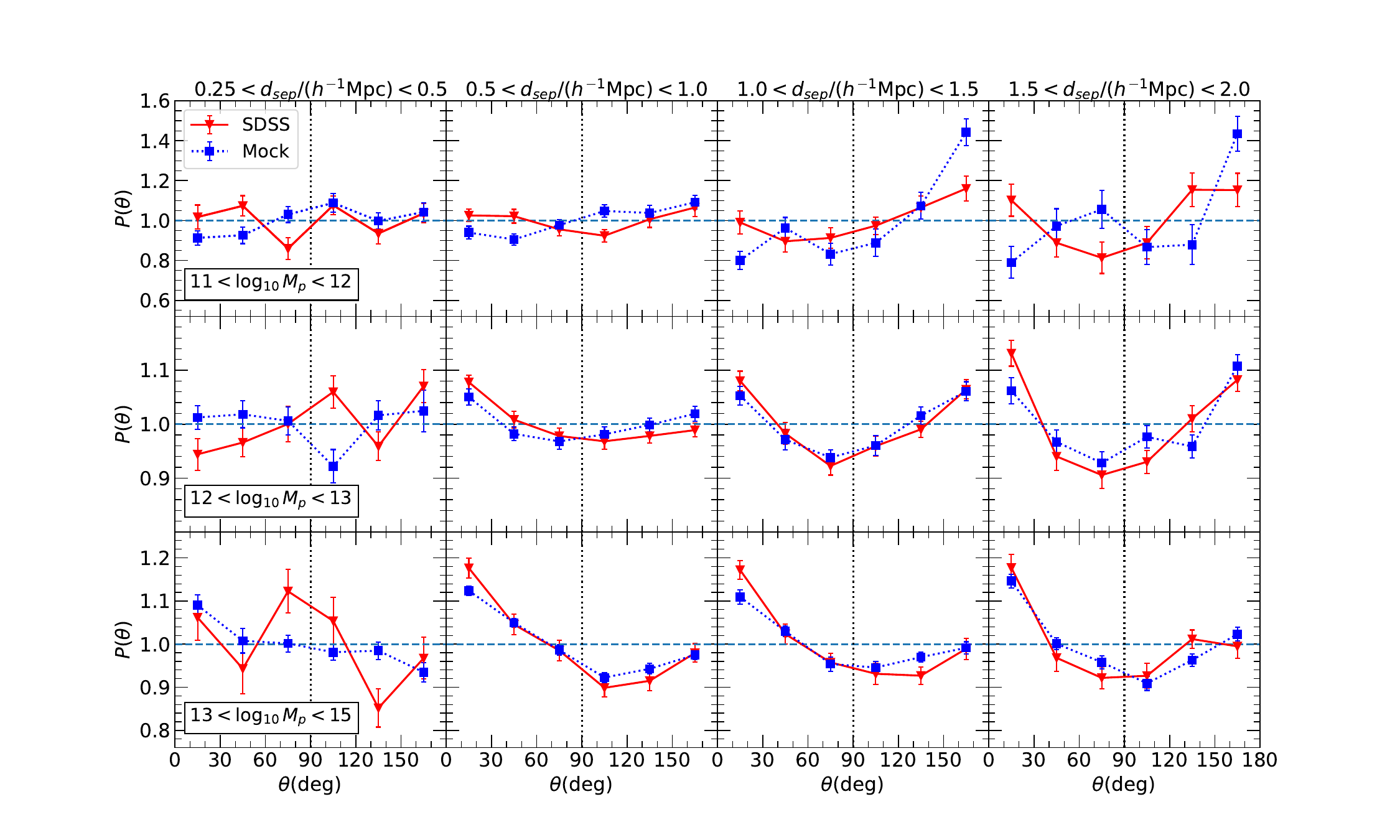}
  \caption{Angular distribution of satellites around halo pairs of different separations and primary halo masses, as indicated. Red and blue symbols/lines are for the SDSS sample and the mock catalog, respectively.}
  \label{fig:Ang_sdss_mock}
\end{figure*}

\begin{figure*}[ht!]
  \centering
  \includegraphics[width=0.9\linewidth]{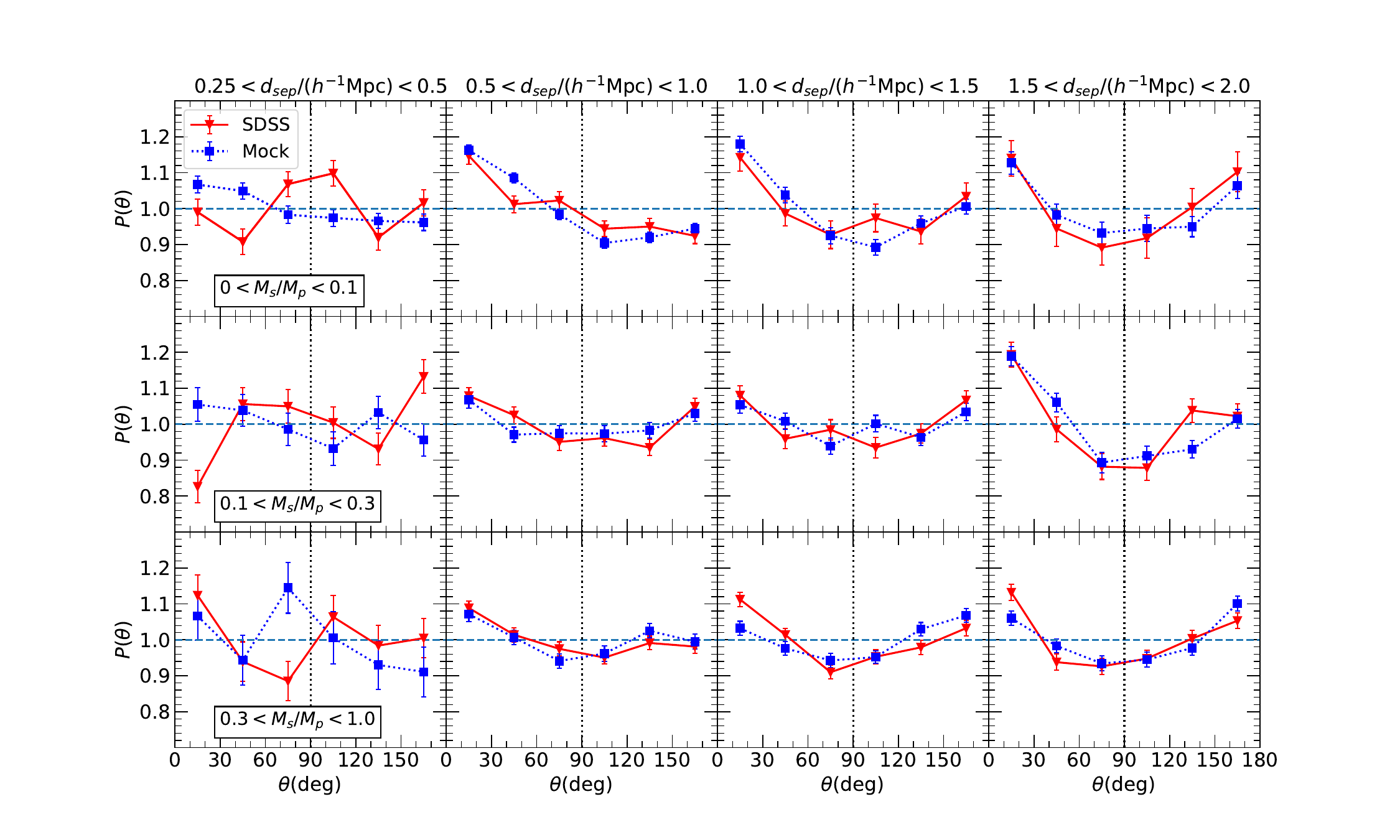}
  \caption{Angular distribution of satellites around halo pairs of different separations and secondary-to-primary halo mass ratios, as indicated. Red and blue symbols/lines are for the SDSS sample and the mock catalog, respectively.}
  \label{fig:Ang_sdss_mock_rt}  
\end{figure*}

\begin{figure*}[ht!]
  \centering 
  \includegraphics[width=0.9\linewidth]{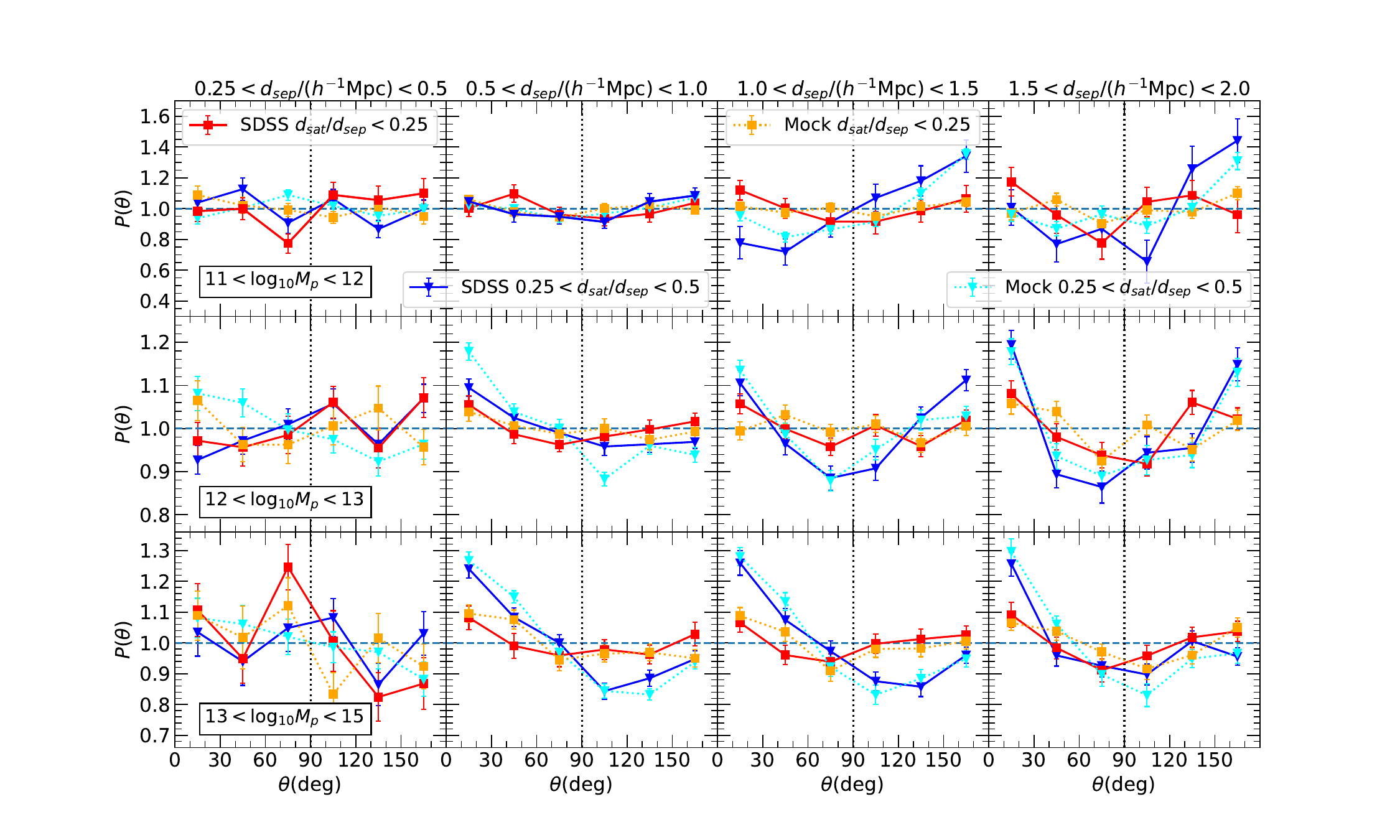}  
  \caption{Angular distribution of satellites around halo pairs of different separations and satellite-to-host distances, as indicated. Red/blue and yellow/cyan symbols/lines are for the SDSS sample and the mock catalog, respectively.}
  \label{fig:Ang_sdss_mock_dsatdsep}
\end{figure*}

We utilize the projected alignment cross-correlation function \citep[ACCF;][]{Faltenbacher2009, Li2013}, denoted as $w_p(r_p,\theta_p)$, to validate our orientation model. In contrast to the traditional projected cross-correlation function (CCF), $w_p(r_p)$, which measures galaxy clustering as a function of the projected separation ($r_p$) between galaxies of given properties and a reference sample of galaxies, the ACCF measures the clustering as a function of both $r_p$ and $\theta_p$, the angle between the major axis of the galaxy in consideration and the connection line to the companion galaxy in the reference sample. The alignment between the orientation of the galaxies and the distribution of reference galaxies is quantified by comparing $w_p(r_p,\theta_p)$ at different values of $\theta_p$. In~\autoref{fig:test_mock_theta}, we present the ratios of $w_p(r_p,\theta_p)$ measured at three non-overlapping intervals of $\theta_p$ with respect to the traditional CCF $w_p(r_p)$. The panels from left to right correspond to different stellar mass bins. Different colors indicate various $\theta_{p}$ bins, with solid lines representing SDSS data and dotted lines representing the mock results. For error estimation, we randomly shuffle the position angles of galaxies and repeat the analysis 20 times. The $1\sigma$ variance of these 20 shuffled samples is used to estimate the errors of the correlation function ratios, as plotted in the figure. As can be seen, the measurements of the alignment signal in the mock catalog are in good agreement with those in the SDSS sample across all mass bins and all scales probed. This result demonstrates that our halo-based empirical model can effectively reproduce the spatial distribution and alignment of galaxies in the local Universe.

\subsection{Comparison between model and data}

To provide a direct comparison with observational data, we employ the same method described in \ref{subsec:data} to identify central galaxies, halo pairs, and satellites in the mock catalog. This process results in 209,935 halo pairs and 331,980 satellites. We then measure the angular distribution of satellites around the paired centrals for different primary halo masses ($M_p$), pair separations ($d_\text{sep}$), secondary-to-primary halo mass ratios ($M_s/M_p$), as well as the distance from satellites to their halos scaled by pair separation ($d_{\text{sat}}/d_{\text{sep}}$). The results are shown in~\autoref{fig:Ang_sdss_mock}, ~\autoref{fig:Ang_sdss_mock_rt}, and~\autoref{fig:Ang_sdss_mock_dsatdsep}, where the corresponding results from SDSS are displayed for comparison. As can be seen, the results from the observational data are reasonably reproduced by the mock catalog across all cases considered, although subtle discrepancies are noted in instances where the measurements are relatively noisy, either from the data, the mock catalog, or both. These results suggest that the $\Lambda$CDM model can naturally reproduce the anisotropic distribution of satellite galaxies in and around pairs of dark matter halos as observed in the local Universe. 

\section{Overlap Effect}
\label{sec:overlap}

In this section, we analyze the overlap effect, which occurs when an overabundance of satellites is naturally found between two halos as they are brought closer together. By comparing the distribution of satellites around these overlapping halos with those around actual halo pairs, we aim to test the hypothesis that the observed lopsided and bulging distribution of satellites might be somewhat attributed to the overlap effect. In~\autoref{sec:overlap_noalign}, we begin by following previous studies to construct a sample of overlapping halos without considering their spatial alignment. We then take into account the alignment of halos in~\autoref{sec:4.2}.

\subsection{Overlap effect with no alignment}
\label{sec:overlap_noalign}

For each central galaxy in the halo pairs, we randomly select a halo from the full sample, ensuring that they closely match in halo mass, within a tolerance of $\mathrm{\Delta log_{10}}(M/[h^{-1}\rm{M_{\odot}}]) < 0.1$. For a given pair of halos, the two selected halos, along with their satellites within a projected distance of $d_{\text{sep}}$ and a line-of-sight velocity separation of $v = 1000~\mathrm{km/s}$, are placed at the same angular position and redshift as the original halo pair. To reduce shot noise, this process is repeated 10 times, resulting in a sample of overlapping halos that is ten times the size of the original halo pair sample. We then measure the angular distribution of the satellites around these overlapping pairs using the same method as before. 

\begin{figure*}[ht!]
  \centering
  \includegraphics[width=0.9\linewidth]{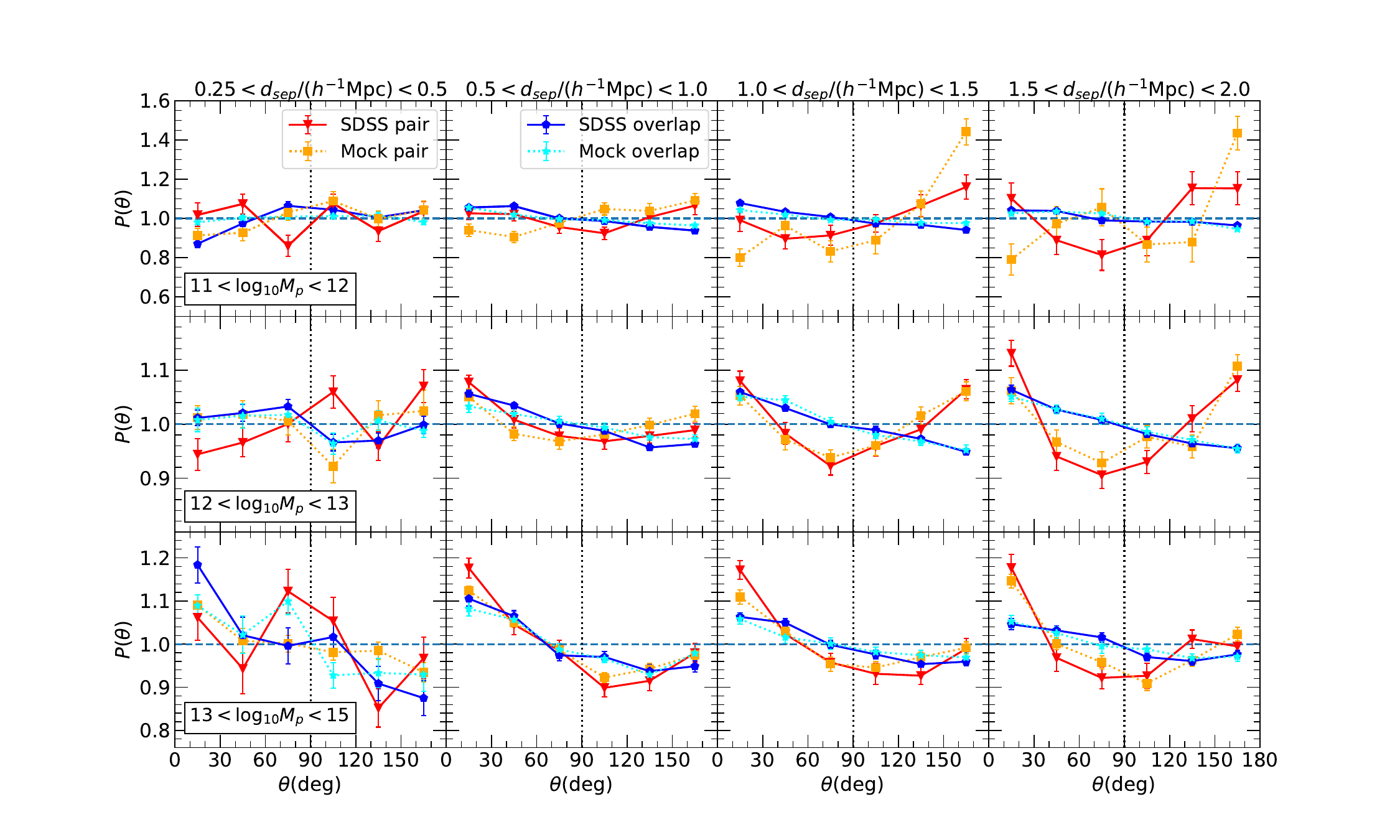}
  \caption{Comparison of satellite distribution between actual halo pairs and the overlap sample, for both the  SDSS (red/blue lines) and the mock catalog (yellow/cyan lines).}
  \label{fig:Ang_sdss_mock_real_norot}
\end{figure*}

\begin{figure*}[ht!]
  \centering
  \includegraphics[width=0.9\linewidth]{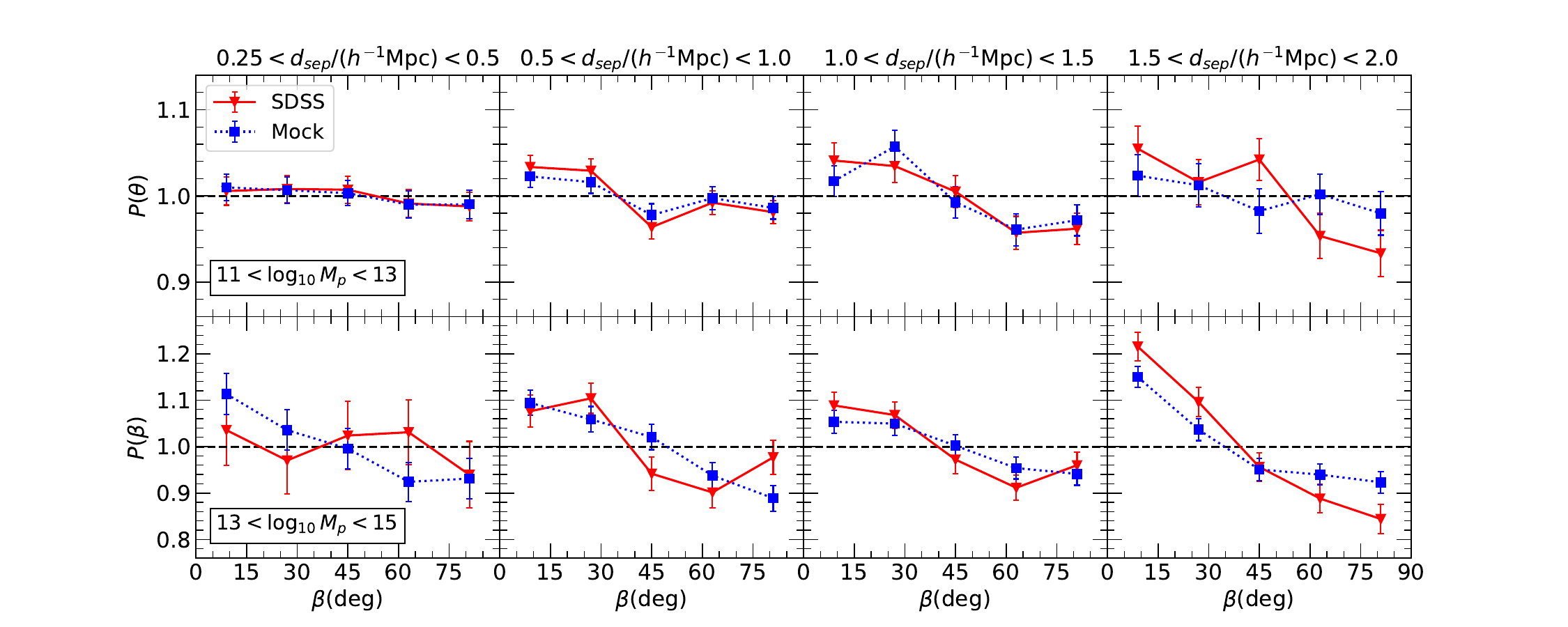}
  \includegraphics[width=0.9\linewidth]{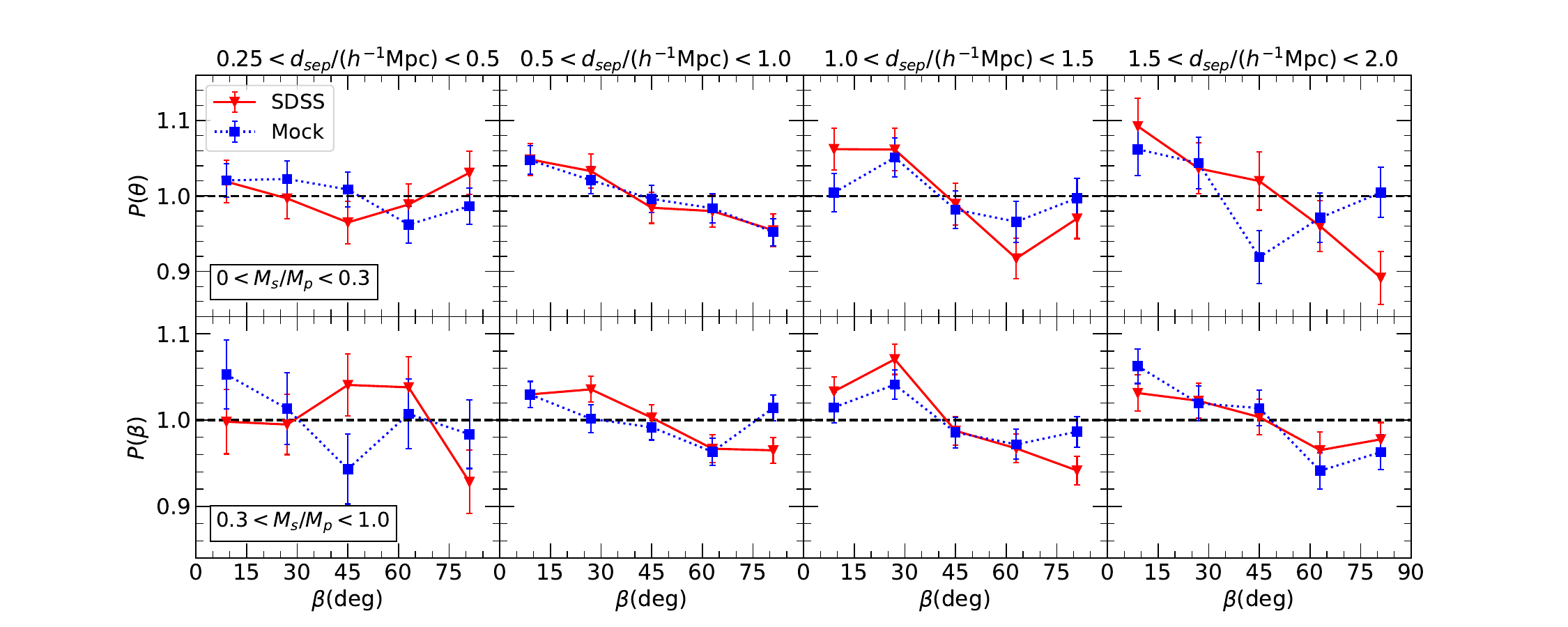}
  \caption{Angular distribution of the central galaxy orientation for primary halos of the pairs, as measured for actual pairs from SDSS (red) and those from the mock catalog (blue). Ranges of pair separation, primary halo mass and secondary-to-primary halo mass ratio are indicated in each panel.}
  \label{fig:beta}
\end{figure*}

We perform this analysis for both the real data from SDSS and the mock catalog. In~\autoref{fig:Ang_sdss_mock_real_norot}, we present the distribution of satellites around the overlapping pairs, where blue represents the SDSS sample and cyan represents the mock catalog, for different halo masses and pair separations. These are compared with the results from actual halo pairs, with red indicating the SDSS sample and orange for the mock catalog. Consistent with previous studies \citep[e.g.][]{Gong2019}, generally, the overlapping pairs show no bulging signal at all, exhibiting only a weakly lopsided distribution characterized by higher $P(\theta)$ at smaller angles $\theta$. This lopsidedness effect is weaker than that observed from the actual pairs, particularly at lower halo masses and larger pair separations. 

\subsection{Overlap effect with alignment} 
\label{sec:4.2}

Now we take into account the alignment of central galaxies and halo pairs in our examination of the overlap effect. Specifically, we consider the alignment between the orientation of the central galaxy of the primary halos and the halo pair, defined by $\beta$, the angle between the major axis of the primary central galaxy and the line connecting the two paired centrals. The measurements of this angle are performed on the sky plane for both the SDSS galaxy sample and the mock catalog, and geometric symmetry naturally constrains the range of $\beta$ to be $[0^{\circ}, 90^{\circ}]$. The position angles (PA) of the major axis of SDSS galaxies are measured from their $r$-band images \citep{Lupton2001}. The major axes of mock galaxies are determined using the method described in~\autoref{sec:mock_construction}. 

\autoref{fig:beta} shows the probability distribution of $\beta$ for different primary halo masses and various pair separations. As can be seen, the mock catalog agrees well with the SDSS sample in all panels, demonstrating again that the alignment signal of the real galaxies is well captured in the mock catalog by our orientation model. As expected, $\beta$ is not randomly distributed; rather, smaller angles are preferred, indicating that the orientation of the central galaxies tends to align with the connection line of the halo pairs. This effect is more evident as one considers pairs with larger separations. This result can be attributed to the limited thickness of filaments, because of which it is natural for pairs with greater separations to align more closely with the filaments, which are known to align with the orientation of central galaxies to some degree (\citealt{Zhang2013,Chen2015}). 



\begin{figure*}[ht!]
  \centering
  \includegraphics[width=0.9\linewidth]{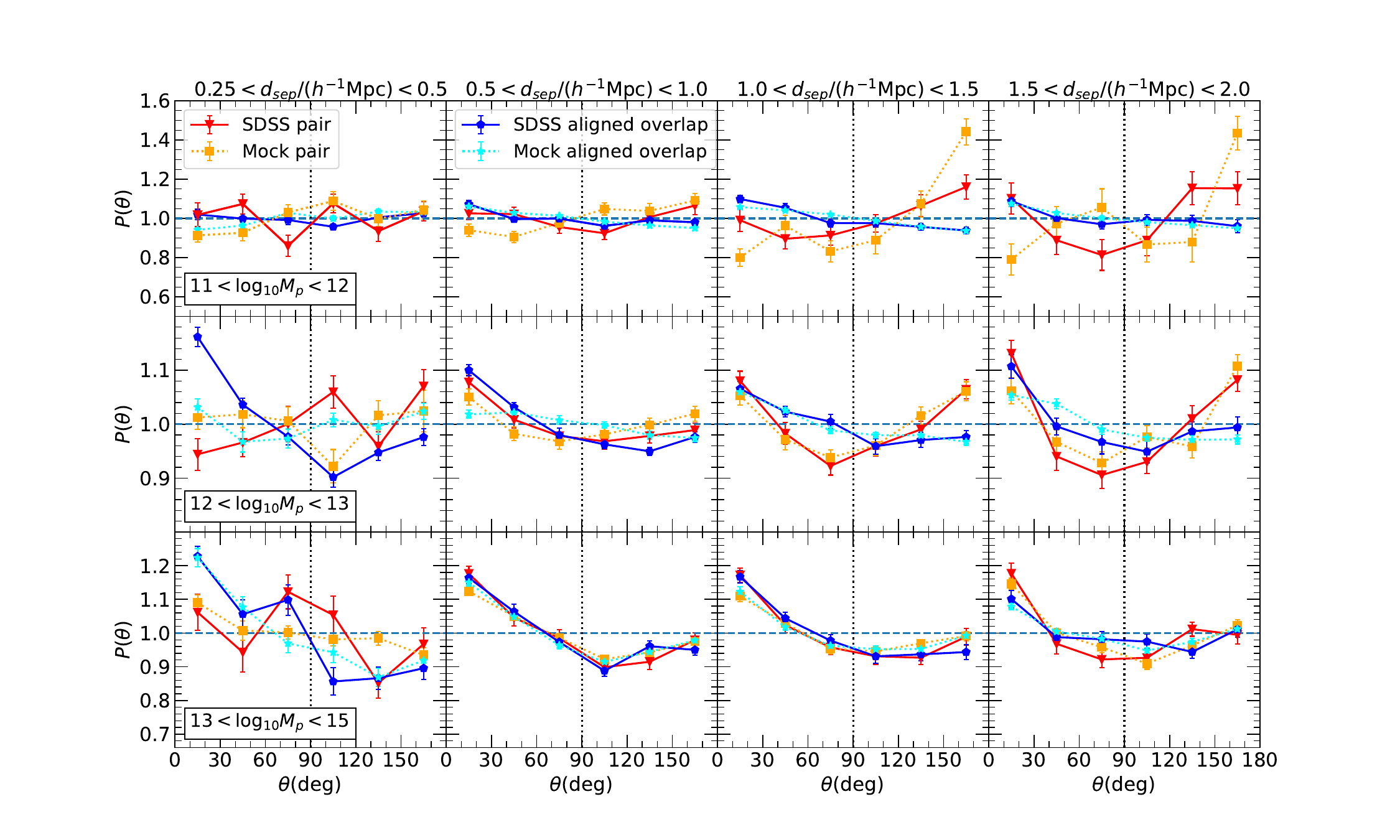}
  \caption{Same as \autoref{fig:Ang_sdss_mock_real_norot}, but the spatial alignment of central galaxies is taken into account when constructing the overlap sample. See the text for details.}
  \label{fig:Ang_sdss_mock_overlap_galaxyaxis_lop}
\end{figure*}

To incorporate the alignment into the overlap effect, we rotate the entire configuration of galaxy distribution around each central galaxy of a given pair in the overlap sample, ensuring that the major axes of the two centrals are aligned with the line connecting the pair. This process is applied consistently for both the SDSS sample and the mock catalog. The samples so constructed are referred to as  ``aligned overlap samples'', in contrast to the sample of randomly placed overlapping halos analyzed above. In \autoref{fig:Ang_sdss_mock_overlap_galaxyaxis_lop}, the angular distribution of satellites in the aligned overlap samples is plotted using blue and cyan symbols/lines for the SDSS and mock catalogs, respectively, across different primary halo mass and pair separation bins. For comparison, the results from the actual halo pairs are plotted using red and orange symbols/lines.

When the alignment is included, the bulging signal manifests, and the lopsidedness strengthens in all cases considered, as anticipated. Consequently, the overlapping halos in the highest halo mass bin ($M_{\text{p}} > 10^{13}~h^{-1} M_\odot$) exhibit behavior that closely resembles that of the actual halo pairs. This holds true for all pair separations in both the SDSS and mock catalog. However, significant discrepancies remain for lower halo masses, particularly at larger pair separations, although better agreements are observed compared to those found for the randomly placed overlapping halo pairs from \autoref{fig:Ang_sdss_mock_real_norot}.

\begin{figure*}[ht!]
  \centering
  \includegraphics[width=0.9\linewidth]{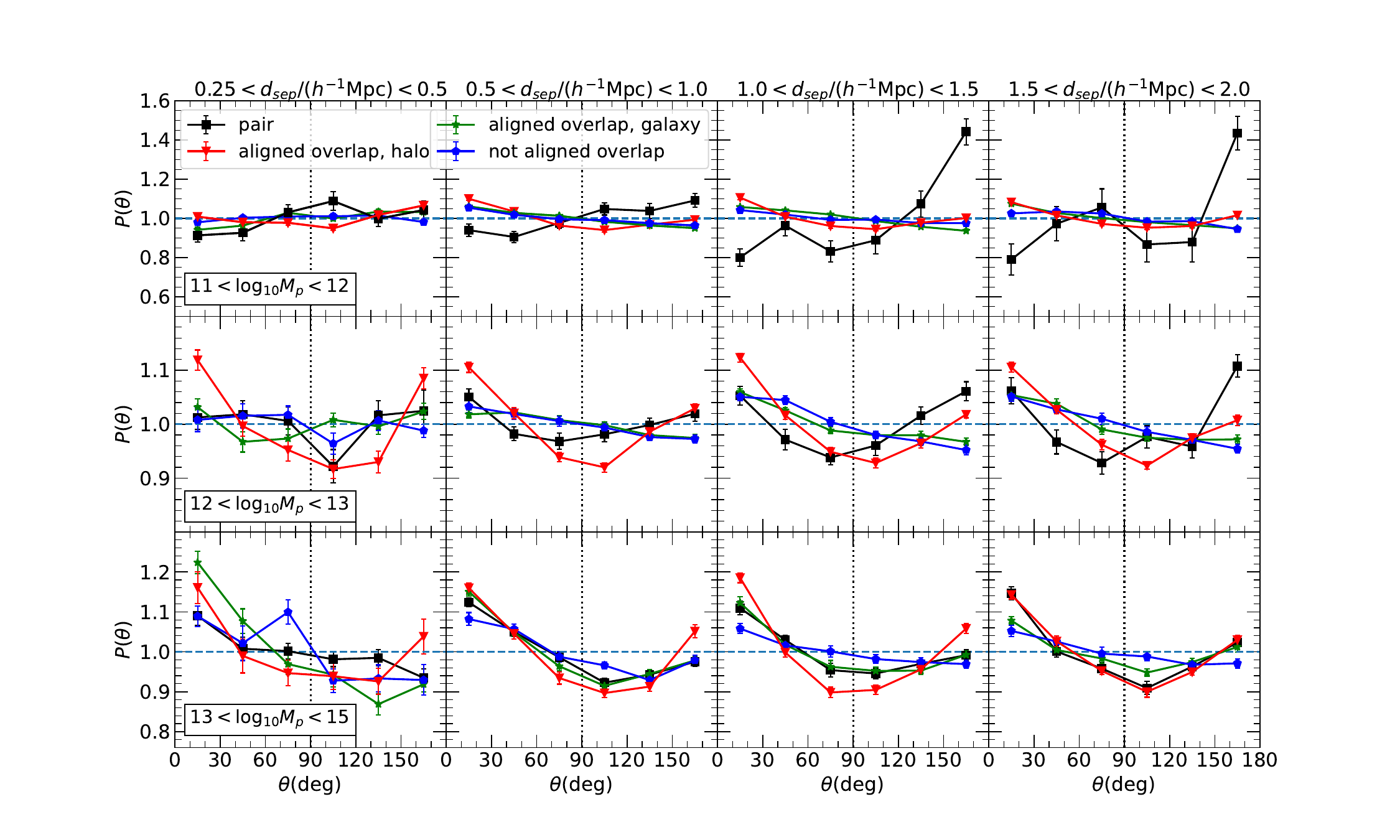}
  \caption{Comparison of the angular distribution of satellites between four different samples constructed from the mock catalog: the actual halo pairs (black), the overlap sample without alignment (blue), and the aligned overlap sample constructed using the galaxy orientation (green) or the halo orientation (red).}
  \label{fig:Ang_mock_overlap}
\end{figure*}

Considering the previous finding that the alignment of halos with large-scale structures is weaker for lower masses of halos \citep[e.g.][]{Faltenbacher2007,Faltenbacher2008}, the discrepancies observed at relatively low masses could be partly attributed to the misalignment between the orientations of central galaxies and those of their host halos. This misalignment has been incorporated into our galaxy orientation model, which has demonstrated good agreement with SDSS across all instances thus far. To test this conjecture, we reconstruct the aligned overlap samples by aligning the host halo orientation with the halo pair connection line, specifically using the major axis of the host halos instead of that of the central galaxies. Since the orientations of halos are not available for the observed galaxies, this analysis is performed solely for the mock catalog. The results are presented in ~\autoref{fig:Ang_mock_overlap} as red symbols/lines, compared with mock catalog measurements for the actual halo pairs (black), the overlapping halos without alignment (blue), and the overlapping halos aligned by galaxy major axes (green). 

As expected, the lopsidedness and bulging signals become stronger when the overlapping halos are aligned by halo orientation instead of galaxy orientation. Consequently, the results for the actual halo pairs can be better explained by the overlap effect at $M_{\text{p}} < 10^{13}~ h^{-1} \text{M}_{\odot}$, although this comes at the expense of slightly worse agreement for more massive halos, for which the overlapping halos aligned with galaxy orientation provide better agreement. This test confirms that the alignment of galaxies or host halos plays an important role in producing the anisotropic distribution of satellites around halo pairs. 

We note that, as can be seen from~\autoref{fig:Ang_sdss_mock_overlap_galaxyaxis_lop} and~\autoref{fig:Ang_mock_overlap}, significant discrepancies still exist between all the overlap samples and the actual halo pairs for $M_{\text{p}} < 10^{12}~h^{-1} \text{M}{\odot}$ and $d{\text{sep}} \gtrsim 1~h^{-1} \text{Mpc}$. In the parallel paper (\citealt{ma2025}), we show that these discrepancies are attributed to the 3D-to-2D projection which affects the satellite distribution for actual halo pairs and overlapping halos in different ways. In general, projection suppresses the bulging signal in the satellite distribution for both actual pairs and overlapping halos, at a given pair separation, halo mass, and mass ratio. However, the strength of this projection effect depends on pair separation differently for actual pairs and overlapping halos. This difference arises because the bulging signal increases with pair separation for actual halo pairs, but decreases with pair separation for overlapping halos. Since a pair identified in 2D corresponds to a pair in 3D with a larger separation, projection thereby enhances the bulging signal for actual pairs (thus reducing the apparent projection effect), while it further diminishes the bulging signal for overlapping halos (thus exacerbating the projection effect). These divergent projection effects result in significant discrepancies between actual pairs and overlapping halos, particularly at low masses and large separations. In fact, in 3D, the satellite distribution of low-mass, widely-separated halo pairs is well reproduced by the overlap sample. For further details, the reader is referred to Section 3.2 and Fig. 9 of that paper. 

 

\section{Summary}
\label{sec:sum}

In this paper, we examine the anisotropic distribution of satellite galaxies in and around pairs of dark matter halos. We first utilize an observational sample based on SDSS DR7, selecting pairs of central galaxies with a wide range of pair separations ($0.25~h^{-1} \text{Mpc} < d_{\text{sep}} < 2~h^{-1} \text{Mpc}$) and primary halo masses ($10^{11}~h^{-1} M_\odot < M_p < 10^{15}~h^{-1} M_\odot$). We measure the angular distribution of satellites for halo pairs of different masses, separations, mass ratios, and satellite-to-central distances. Next, we construct a mock catalog based on an empirical model of galaxy formation, which is implemented in the Illustris-TNG300 simulation, to facilitate direct comparisons between the model and the data regarding satellite distribution. Finally, we investigate the overlap effect by constructing samples of overlapping halos in both the SDSS and the mock catalog. In particular, we consider the spatial alignment of halos in the overlap effect, which has been completely overlooked in previous studies.

The main results of this paper can be summarized as follows.

\begin{enumerate}[label=(\roman*)]
    \item The angular distribution of satellite galaxies in pairs of dark matter halos exhibits a pronounced tendency toward lopsidedness, with satellites preferentially located between the two central galaxies. Additionally, there is a significant bulging distribution, characterized by a higher concentration of satellites along the line connecting the two centrals compared to those found perpendicular to it. Both the lopsided and bulging distributions strengthen as pair separation and primary halo mass increase; however, there is no significant dependence on the mass ratio of halo pairs. Satellites located \textbf{further}  to their centrals demonstrate stronger lopsided and bulging distributions than those that are \textbf{closer}.
    \item The mock catalog successfully reproduces the observational measurements of the angular distribution of satellites around pairs of dark matter halos. It also accurately models the spatial alignment of galaxies with varying masses and the alignment between the orientation of central galaxies and the orientation of halo pairs.
    \item The lopsided and bulging distribution of satellites observed in both the SDSS and the mock catalog can largely be explained by overlapping two randomly selected halos of similar masses along with their surrounding satellite distribution, provided that the alignment between the orientations of the halos and the line connecting the halo pairs is considered. This finding suggests that the angular distribution of satellites around halo pairs can be naturally explained by the alignment of halos and galaxies with respect to large-scale filamentary structures in a $\Lambda$CDM universe.
\end{enumerate}

We should point out that different definitions of halo (galaxy) pairs have been adopted in previous studies. For instance, \cite{Libeskind2016} identified galaxy groups consisting of only two galaxies to constitute the sample of galaxy pairs, while \cite{Pawlowski2017} required galaxy pairs to be the two most luminous galaxies within $1.5~\rm{Mpc}$ of the pair’s midpoint. In this work, we follow the selection criteria of \citet{Gong2019}, identifying halo pairs by the two most massive galaxies within a radius equal to the pair separation from the pair’s midpoint. Compared to \citet{Gong2019}, we have extended the range of halo pair separations and halo masses. \citet{Gong2019} focused mainly on halo pairs similar to the Local Group, while we aim to extend their analysis to more general situations of halo pairs in the local universe. Nevertheless, our sample also includes halo pairs that resemble the Local Group.

Unlike our results, \citet{Gong2019} found that the lopsided signal is strongest for the least massive hosts and the most extreme mass ratios. This discrepancy could be attributed to the projection effects and selection effects in the observational sample, as \citet{Gong2019} conducted their analysis purely in simulation. In a parallel paper (\citealt{ma2025}), we examine the satellite distribution around halo pairs in three-dimensional space using the Illustris-TNG300 simulation and compare the results with those obtained in both two-dimensional space and a SDSS-like redshift survey. Our 3D results are consistent with \citet{Gong2019}. 

The impacts of projection and selection effects on the measurements of satellite distribution are discussed in detail in \cite{ma2025}. In short, after statistical correction, selection effects in the observed sample introduce negligible biases in satellite distribution measurements, but the 3D-to-2D projection significantly suppresses bulging signals, particularly for widely separated, low-mass halo pairs. Consequently, this projection effect transforms the anti-correlation between bulging signal and halo mass observed in 3D into a positive correlation in observational data. As noted in \cite{ma2025}, this projection effect cannot be straightforwardly corrected, and a robust understanding requires careful comparisons between models and observational data using mock catalogs.



Regarding the origin of the anisotropic distribution of satellites in halo pairs, our conclusion aligns with the findings of \cite{Pawlowski2017} and \citet{Gong2019}. While \citet{Gong2019} traced the trajectories of satellites in simulation, we artificially included the filamentary structures around halo pairs in our overlap samples. The influence of filaments on the anisotropic signal can be further verified by combining the trajectories of satellites with the cosmic web in which they reside. Our analysis demonstrates that the overlap effect, when combined with halo alignment, provides a comprehensive explanation for both the lopsided and bulging satellite distributions. The overlap effect arises naturally from the overabundance of satellites in and around close pairs, combined with the alignment among galaxies, dark matter halos, and large-scale structures, which have been well established across a broad range of spatial scales and can be effectively reproduced by halo-based models implemented in a $\Lambda$CDM cosmological simulation (\citealt{Yang2006, Faltenbacher2008, Faltenbacher2009, Li2013}, among others). 

Our work suggests that alternative mechanisms, such as the local gravitational effects proposed by \citet{Libeskind2016}, may not be required to explain these particular observational trends. In fact, previous investigations considered such alternatives primarily because models invoking overlap without alignment failed. Our work shows that incorporating alignment resolves these failures, making the overlap effect a sufficient explanation in most cases and thereby limiting the scope where additional mechanisms might be needed. Nevertheless, these alternative mechanisms cannot be definitively ruled out by our study; direct tests would require focused future work. As we discuss in more detail in our parallel paper (\citealt{ma2025}), the origin of the satellite distribution of halo pairs is more clearly revealed in three dimensions, and a correct understanding of the picture can be achieved only by carefully comparing the model and data with the help of mock catalogs. 




\section*{Acknowledgments}
This work is supported by the National Key R\&D Program of China (grant NO. 2022YFA1602902), the National Natural Science Foundation of China (grant Nos. 12433003, 11821303, 11973030), and the China Manned Space Program with grant no. CMS-CSST-2025-A10.

Funding for SDSS and SDSS-II has been provided by the
Alfred P. Sloan Foundation, the Participating Institutions, the
National Science Foundation, the U.S. Department of Energy,
the National Aeronautics and Space Administration, the
Japanese Monbukagakusho, the Max Planck Society, and the
Higher Education Funding Council for England. The SDSS
website is (\url{http://www.sdss.org/}). SDSS is managed by the Astrophysical Research Consortium for the Participating Institutions. The Participating
Institutions are the American Museum of Natural History,
Astrophysical Institute Potsdam, University of Basel, University of Cambridge, Case Western Reserve University,
University of Chicago, Drexel University, Fermilab, the
Institute for Advanced Study, the Japan Participation Group,
Johns Hopkins University, the Joint Institute for Nuclear
Astrophysics, the Kavli Institute for Particle Astrophysics and
Cosmology, the Korean Scientist Group, the Chinese Academy
of Sciences (LAMOST), Los Alamos National Laboratory, the
Max-Planck-Institute for Astronomy (MPIA), the Max-PlanckInstitute for Astrophysics (MPA), New Mexico State University, Ohio State University, University of Pittsburgh,
University of Portsmouth, Princeton University, the United
States Naval Observatory, and the University of Washington.

%

\vspace{5mm}

\bibliography{sample631}{}
\bibliographystyle{aasjournal}



\end{document}